\documentclass{article}
\usepackage{amssymb,amsmath,amsfonts,graphicx,hyperref}
\input{tcilatex}

\newcommand{\Y}{Y\to X}
\newcommand{\la}{\lambda}
\newcommand{\al}{\alpha}
\newcommand{\bt}{\beta}
\newcommand{\w}{\wedge}
\newcommand{\cO}{{\bf\Omega}}
\newcommand{\m}{\mu}
\newcommand{\dr}{\partial}
\newcommand{\bL}{{\cal L}}
\newcommand{\wh}{\widehat}
\newcommand{\G}{\Gamma}
\newcommand{\beq}{\begin{equation}}
\newcommand{\eeq}{\end{equation}}
\newcommand{\be}{\begin{eqnarray*}}
\newcommand{\ee}{\end{eqnarray*}}

\def\op#1{\mathop{\fam0 #1}\limits}

\begin{document}

\title{Jet Methods in Time--Dependent Lagrangian Biomechanics}\author{Tijana T. Ivancevic\\ {\small Society for Nonlinear Dynamics in Human Factors, Adelaide, Australia}\\
{\small and}\\
{\small CITECH Research IP Pty Ltd, Adelaide, Australia}\\
{\small e-mail: ~tijana.ivancevic@alumni.adelaide.edu.au}}\date{}\maketitle

\begin{abstract}
In this paper we propose the time-dependent generalization of an `ordinary' autonomous human biomechanics, in which \emph{total mechanical + biochemical energy is not conserved}. We introduce a general framework for time-dependent biomechanics in terms of jet manifolds associated to the extended musculo-skeletal configuration manifold, called the configuration bundle. We start with an ordinary configuration manifold of human body motion, given as a set of its all active degrees of freedom (DOF) for a particular movement. This is a Riemannian manifold with a material metric tensor given by the total mass-inertia matrix of the human body segments. This is the base manifold for standard autonomous biomechanics. To make its time-dependent generalization, we need to extend it with a real time axis. By this extension, using techniques from fibre bundles, we defined the biomechanical configuration bundle. On the biomechanical bundle we define vector-fields, differential forms and affine connections, as well as the associated jet manifolds. Using the formalism of jet manifolds of velocities and accelerations, we develop the time-dependent Lagrangian biomechanics. Its underlying geometric evolution is given by the Ricci flow equation.\\

\noindent\textbf{Keywords:} Human time-dependent biomechanics, configuration bundle, jet spaces, Ricci flow
\end{abstract}

%\tableofcontents

\section{Introduction}

It is a well-known fact that most of dynamics in both classical and quantum physics is based on \emph{assumption of a total energy conservation} (see, e.g. \cite{GaneshADG}). Dynamics based on this assumption of time-independent energy, usually given by Lagrangian or Hamiltonian energy function, is called \emph{autonomous}. This basic assumption is naturally inherited in human biomechanics, formally developed using Newton--Euler, Lagrangian or Hamiltonian formalisms (see \cite{GaneshSprSml,GaneshWSc,GaneshSprBig,StrAttr,TijIJHR,TijNis,TijNL,TijSpr}).

And this works fine for most individual movement simulations and predictions, in which the total human energy dissipations are insignificant. However, if we analyze a 100\,m-dash sprinting motion, which is in case of top athletes finished under 10\,s, we can recognize a significant slow-down after about 70\,m in \emph{all} athletes -- despite of their strong intention to finish and win the race, which is an obvious sign of the total energy dissipation. This can be
seen, for example, in a current record-braking speed–distance curve of Usain Bolt, the  world-record holder with 9.69 s \cite{SciSport}, or in a former record-braking speed–distance curve of Carl Lewis, the former world-record holder (and 9 time Olympic gold medalist) with 9.86 s (see Figure 3.7 in \cite{TijSpr}). In other words, the \emph{total mechanical + biochemical energy} of a sprinter \emph{cannot be conserved} even for 10\,s. So, if we want to develop a realistic model of intensive human motion that is longer than 7--8\,s (not to speak for instance of a 4 hour tennis match), we necessarily need to use the more advanced formalism of time-dependent mechanics.

Similarly, if we analyze individual movements of gymnasts, we can clearly see that the high speed of these movements is based on quickly-varying mass-inertia distribution of various body segments (mostly arms and legs). Similar is the case of pirouettes in ice skating. As the total mass-inertia matrix $M_{ij}$ of a biomechanical system corresponds to the Riemannian metric tensor $g_{ij}$ of its configuration manifold, we can formulate this problem in terms of time-dependent Riemannian geometry \cite{GaneshSprBig,GaneshADG}.

The purpose of this paper is to introduce a general framework for time-dependent biomechanics, consisting of the following four steps:
\begin{enumerate}
  \item Human biomechanical configuration manifold and its (co)tangent bundles;
  \item Biomechanical configuration bundle, as a time--extension of the configuration manifold;
  \item Biomechanical jet spaces and prolongation of locomotion vector-fields developed on the configuration bundle; and
  \item Time--dependent Lagrangian dynamics using biomechanical jet spaces.
\end{enumerate}
In addition, we will show that Riemannian geometrical basis of this framework is defined by the Ricci flow. In particular, we will show that the exponential--like decay of total biomechanical energy (due to exhaustion of biochemical resources \cite{TijSpr}) is closely related to the Ricci flow on the configuration manifold of human motion.

\section{Configuration Manifold for Autonomous Biomechanics}

Recall from \cite{TijIJHR} that representation of an ideal humanoid--robot motion is rigorously defined in
terms of {rotational} constrained $SO(3)$--groups in all main robot joints. Therefore, the {configuration manifold}
$Q_{rob}$
for humanoid dynamics is defined as a topological product of all included $%
SO(3)$ groups, $Q_{rob}=\prod_{i}SO(3)^{i}$. Consequently, the
natural stage for autonomous Lagrangian dynamics of robot motion
is the {tangent
bundle} $TQ_{rob}$, defined as follows.
To each $n-$dimensional ($n$D) {configuration
manifold} $Q$ there is associated its $2n$D {\it velocity phase--space
manifold}, denoted by $TQ$ and called the tangent bundle of $Q$.
The original smooth manifold $Q$ is called the {\it base} of $TQ$.
There is an onto map $\pi :TQ\rightarrow Q$, called the
{\it projection}. Above each point $x\in Q$ there is a {tangent space}
$T_{x}Q=\pi ^{-1}(x)$ to $Q$ at $x$, which
is called a {fibre}. The fibre $T_{x}Q\subset TQ$ is the subset of $%
TQ $, such that the total tangent bundle,
$TQ=\dbigsqcup\limits_{m\in Q}T_{x}Q$, is a {disjoint union} of
tangent spaces $T_{x}Q$ to $Q$ for all points $x\in Q$. From
dynamical perspective, the most important quantity in the tangent
bundle concept is the smooth map $v:Q\rightarrow TQ$, which
is an inverse to the projection $\pi $, i.e, $\pi \circ v=\func{Id}%
_{Q},\;\pi (v(x))=x$. It is called the {\it velocity vector--field}.
Its graph $(x,v(x))$ represents the {cross--section} of the
tangent
bundle $TQ$. This explains the dynamical term {\it velocity phase--space}, given to the tangent bundle $TQ$ of the manifold $Q$. The tangent bundle
is where tangent vectors live, and is itself a smooth manifold.
Vector--fields are cross-sections of the tangent bundle.
Robot's \emph{Lagrangian} (energy function) is a natural energy
function on the tangent bundle $TQ$.\footnote{
The corresponding autonomous
Hamiltonian robot dynamics takes place in the {cotangent bundle} $T^{\ast }Q_{rob}$, defined as follows.
A {dual} notion to the tangent space $T_{m}Q$ to a smooth manifold
$Q$ at a point $m$ is its {cotangent space} $T_{m}^{\ast }Q$ at
the same point $m$. Similarly to the tangent bundle, for a smooth
manifold $Q$ of dimension $n$, its {cotangent bundle} $T^{\ast }Q$
is the disjoint union of all its cotangent spaces $T_{m}^{\ast }Q$
at all points $m\in Q$, i.e., $T^{\ast }Q=\dbigsqcup\limits_{m\in
Q}T_{m}^{\ast }Q$. Therefore, the cotangent bundle of an
$n-$manifold $Q$ is the vector bundle $T^{\ast }Q=(TQ)^{\ast }$,
the (real) dual of the tangent bundle $TQ$. The cotangent bundle
is where 1--forms live, and is itself a smooth manifold.
Covector--fields (1--forms) are cross-sections of the cotangent
bundle. Robot's \emph{Hamiltonian} is a natural energy function on the
cotangent bundle.}

\begin{figure}[tbh]
\centering \includegraphics[width=13cm]{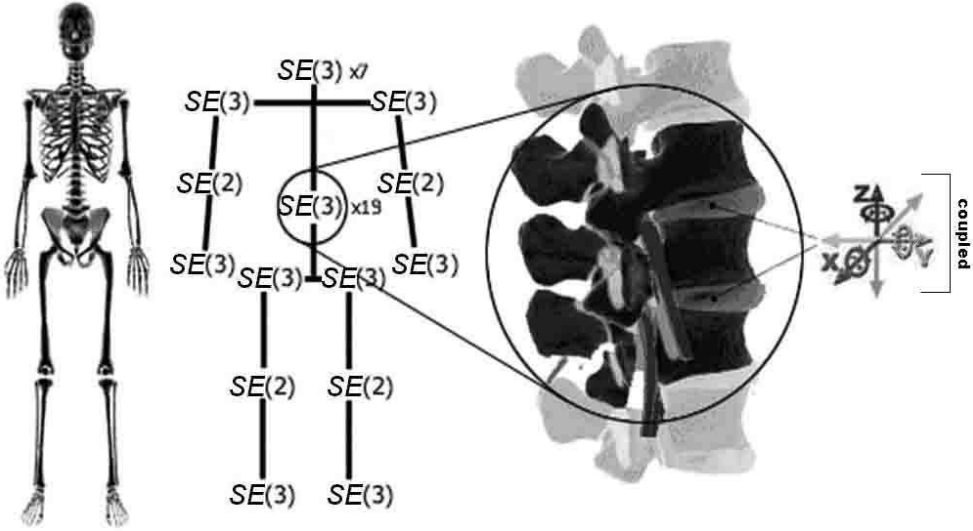} \caption{The
configuration manifold $Q$ of the human musculo-skeletal dynamics is defined as a
topological product of constrained $SE(3)$ groups acting in all
major (synovial) human joints, $Q=\prod_{i}SE(3)^{i}$.}
\label{SpineSE(3)}
\end{figure}

On the other hand, human joints are more flexible than robot joints. Namely,
every rotation in all synovial human joints is followed by the corresponding
micro--translation, which occurs after the rotational amplitude is reached
\cite{TijIJHR}. So, representation of human motion is rigorously defined
in terms of {Euclidean} $SE(3)$--groups of full rigid--body motion \cite{GaneshSprSml,GaneshSprBig,GaneshADG} in all main human joints (see
Figure \ref{SpineSE(3)}). Therefore, the configuration manifold $Q$
for human dynamics is defined as a topological product of all included
constrained $SE(3)$--groups,\footnote{Briefly, the Euclidean SE(3)--group is defined as a semidirect
(noncommutative) product (denoted by $\rhd$) of 3D rotations and 3D translations: ~$%
SE(3):=SO(3)\rhd \mathbb{R}^{3}$. Its most important subgroups are the
following (for technical details see \cite%
{GaneshSprBig,ParkChung,GaneshADG}):\\
{{\frame{$%
\begin{array}{cc}
\mathbf{Subgroup} & \mathbf{Definition} \\ \hline
\begin{array}{c}
SO(3),\text{ group of rotations} \\
\text{in 3D (a spherical joint)}%
\end{array}
&
\begin{array}{c}
\text{Set of all proper orthogonal } \\
3\times 3-\text{rotational matrices}%
\end{array}
\\ \hline
\begin{array}{c}
SE(2),\text{ special Euclidean group} \\
\text{in 2D (all planar motions)}%
\end{array}
&
\begin{array}{c}
\text{Set of all }3\times 3-\text{matrices:} \\
\left[
\begin{array}{ccc}
\cos \theta & \sin \theta & r_{x} \\
-\sin \theta & \cos \theta & r_{y} \\
0 & 0 & 1%
\end{array}%
\right]%
\end{array}
\\ \hline
\begin{array}{c}
SO(2),\text{ group of rotations in 2D} \\
\text{subgroup of }SE(2)\text{--group} \\
\text{(a revolute joint)}%
\end{array}
&
\begin{array}{c}
\text{Set of all proper orthogonal } \\
2\times 2-\text{rotational matrices} \\
\text{ included in }SE(2)-\text{group}%
\end{array}
\\ \hline
\begin{array}{c}
\mathbb{R}^{3},\text{ group of translations in 3D} \\
\text{(all spatial displacements)}%
\end{array}
& \text{Euclidean 3D vector space}%
\end{array}%
$}}}} $Q=\prod_{i}SE(3)^{i}$. Consequently, the
natural stage for autonomous Lagrangian dynamics of human motion is the
tangent bundle $TQ$ (and for the corresponding
autonomous Hamiltonian dynamics is the cotangent bundle $T^{\ast }Q$).

Therefore, the \emph{configuration manifold} $Q$
for human musculo-skeletal dynamics is defined as a Cartesian product of all included
constrained $SE(3)$ groups, $Q=\prod_{j}SE(3)^{j}$ where $j$ labels the active joints. The configuration manifold $Q$ is coordinated by local joint coordinates $x^i(t),~i=1,...,n=$ total number of active DOF. The corresponding joint velocities $\dot{x}^i(t)$ live in the \emph{velocity phase space} $TQ$, which is the \emph{tangent bundle} of the configuration manifold $Q$.

The velocity phase space $TQ$ has the Riemannian geometry with the \textit{local metric form}: $$\langle g\rangle\equiv ds^{2}=g_{ij}dx^{i}dx^{j},\qquad\text{(Einstein's summation convention is in always use)}$$
where $g_{ij}(x)$ is the material metric tensor defined by the biomechanical system's \emph{mass-inertia matrix} and $dx^{i}$
are differentials of the local joint coordinates $x^i$ on $Q$. Besides giving the local
distances between the points on the manifold
$Q,$ the Riemannian metric form $\langle g\rangle$
defines the system's kinetic energy: $$T=\frac{1}{2}g_{ij}\dot{x}^{i}\dot{x}^{j},$$
giving the \emph{Lagrangian equations} of the conservative skeleton motion with kinetic-minus-potential energy Lagrangian $L=T-V$, with the corresponding \emph{geodesic form} \cite{TijNL}
\begin{equation}
\frac{d}{dt}L_{\dot{x}^{i}}-L_{x^{i}}=0\qquad\text{or}\qquad \ddot{x}^i+\Gamma _{jk}^{i}\dot{x}^{j}\dot{x}^{k}=0, \label{geodes}
\end{equation}%
where subscripts denote partial derivatives, while $\Gamma _{jk}^{i}$ are the Christoffel symbols of
the affine Levi-Civita connection of the biomechanical manifold $Q$.

This is the basic geometrical structure for \emph{autonomous Lagrangian biomechanics}. In the next sections will extend this basic structure to embrace the time-dependent Lagrangian biomechanics.

\section{Biomechanical Bundle, Sections and Connections}

While standard autonomous Lagrangian biomechanics is developed on the configuration manifold $X$, the \emph{time--dependent
biomechanics} necessarily includes also the real time axis $\mathbb{R}$, so we have an \emph{extended configuration manifold} $\mathbb{R}\times X$. Slightly more generally, the fundamental geometrical structure is the so-called \emph{configuration bundle}
$\pi:X\rightarrow \mathbb{R}$. Time-dependent biomechanics is thus formally developed either on the \emph{extended configuration manifold} $\mathbb{R}\times X$, or on the configuration bundle $\pi:X\rightarrow \mathbb{R}$, using the concept of \textit{jets}, which are based on the idea of \textit{higher--order tangency}, or higher--order
contact, at some designated point (i.e., certain joint) on a biomechanical configuration manifold $X$.

\begin{figure}[h]
 \centerline{\includegraphics[width=11cm]{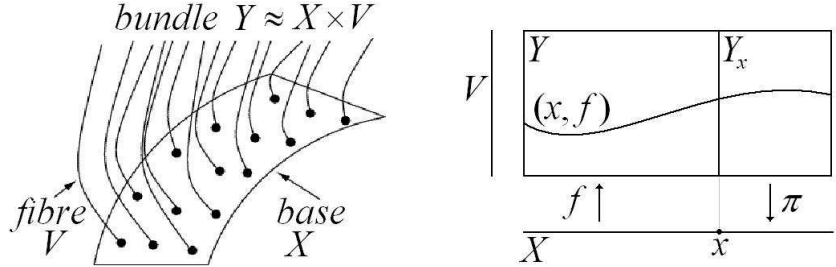}}
\caption{A sketch of a locally trivial fibre bundle $Y\approx X\times V$ as a
generalization of a product space $X\times V$; left -- main
components; right -- a few details (see text for
explanation).}\label{Fibre1}
\end{figure}

In general, tangent and cotangent bundles, $TM$ and $T^{\ast }M$, of a smooth manifold $M$, are special cases of a more general geometrical object called
\emph{fibre bundle}, denoted $\pi
:Y\rightarrow X$, where the word \emph{fiber} $V$ of a map $\pi
:Y\rightarrow X$ is the \emph{preimage} $\pi^{-1}(x)$ of an
element $x\in X$. It is a space which \emph{locally} looks like a
product of two spaces (similarly as a manifold locally looks like
Euclidean space), but may possess a different \emph{global}
structure. To get a visual intuition behind this fundamental
geometrical concept, we can say that a fibre bundle $Y$ is a
\emph{homeomorphic generalization} of a \emph{product space}
$X\times V$ (see Figure \ref{Fibre1}), where $X$ and $V$ are
called the \emph{base} and the \emph{fibre}, respectively. $\pi
:Y\rightarrow X$ is called the \emph{projection}, $Y_{x}=\pi
^{-1}(x)$ denotes a fibre over a point $x$ of the base $X$, while
the map $f=\pi ^{-1}:X\rightarrow Y$ defines the
\emph{cross--section}, producing the \textit{graph} $(x,f(x))$ in
the bundle $Y$ (e.g., in case of a tangent bundle, $f=\dot{x}$
represents a velocity vector--field).

More generally, a biomechanical configuration bundle, $\pi :\Y$, is a locally trivial fibred (or, projection) manifold over the base $X$. It is endowed with an atlas of fibred bundle coordinates $(x^\lambda, y^i)$, where
$(x^\la)$ are coordinates of $X$.

All dynamical objects in time--dependent biomechanics (including vectors, tensors, differential forms and gauge potentials) are \emph{cross--sections} of biomechanical bundles, representing generalizations of
graphs of continuous functions.

An \emph{exterior differential form} $\al$ of order $p$ (or, a $p-$\emph{form} $\al$) on a base manifold $X$ is a section of the bundle $\op\w^p T^*X\to X$ \cite{book}. It has the following expression in local coordinates on $X$
$$
\al =\al_{\la_1\dots\la_p}
dx^{\la_1}\w\cdots\w dx^{\la_p} \qquad ({\rm such~that~~}|\al|=p),
$$
where summation is performed over all ordered collections $(\la_1,...,\la_p)$.
$\cO^p(X)$ is the vector space of
$p-$forms on a biomechanical
manifold $X$. In particular, the 1--forms
are called the {\it Pfaffian forms}.

The \emph{contraction} $\rfloor$ of any vector-field $u = u^\m\dr_\m$ and a $p-$form $\al =\al_{\la_1\dots\la_p}
dx^{\la_1}\w\cdots\w dx^{\la_p}$
on a biomechanical manifold $X$ is given in local coordinates on $X$ by
$$
u\rfloor\al = u^\m
\al_{\m\la_1\ldots\la_{p-1}}
dx^{\la_1}\w\cdots \w dx^{\la_{p-1}}.
$$
It satisfies the following relation
$$
u\rfloor(\al\w\bt)= u\rfloor\al\w\bt +(-1)^{|\al|}\al\w u\rfloor\bt.
$$

The {\it Lie derivative} ${\cal L}_u\al$ of $p-$form $\al$ along
a vector-field $u$ is defined by Cartan's `magic' formula (see \cite{GaneshSprBig,GaneshADG}):
$$\bL_u\al =u\rfloor d\al +d(u\rfloor\al).
$$
It satisfies the \emph{Leibnitz relation}
$$
\bL_u(\al\w\bt)= \bL_u\al\w\bt +\al\w\bL_u\bt.
$$

A linear connection $\bar\G$ on a biomechanical bundle $Y\to X$ is given in local coordinates on $Y$ by \cite{book}
\beq
\bar\G=dx^\la\otimes[\dr_\la-\G^i{}_{j\la}(x)y^j\dr_i]. \label{8}
\eeq

An affine connection $\G$ on a biomechanical bundle $Y\to X$ is given in local coordinates on $Y$ by
\be
\G=dx^\la\otimes[\dr_\la+(-\G^i{}_{j\la}(x)y^j+\G^i{}_\la (x)) \dr_i].
\ee
Clearly, a linear connection $\bar\G$ is a special case of an affine connection $\G$.

\section{Biomechanical Jets}

\begin{figure}[h]
\centerline{\includegraphics[width=6cm]{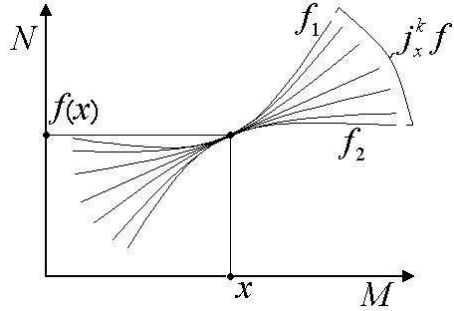}} \caption{An
intuitive geometrical picture behind the $k-$jet concept, based on
the idea of a higher--order tangency (or, higher--order contact). }
\label{jet1}
\end{figure}

A
pair of smooth manifold maps, ~$f_{1},f_{2}:M\rightarrow N$~ (see
Figure \ref{jet1}), are said to be $k-$\emph{tangent} (or
\emph{tangent of order }$k$, or
have a $k$th \emph{order contact}) at a point $x$ on a domain manifold $M$, denoted by $%
f_{1}\sim f_{2}$, iff
\begin{eqnarray*}
f_{1}(x) &=&f_{2}(x)\qquad \text{called}\quad 0-\text{tangent}, \\
\partial _{x}f_{1}(x) &=&\partial _{x}f_{2}(x),\qquad \text{called}\quad 1-%
\text{tangent}, \\
\partial _{xx}f_{1}(x) &=&\partial _{xx}f_{2}(x),\qquad \text{called}\quad 2-%
\text{tangent}, \\
&&...\qquad \text{etc. to the order }k
\end{eqnarray*}
In this way defined $k-$\emph{tangency} is an \emph{equivalence
relation}.

A $k-$\textit{jet} (or, a \emph{jet of order }$k$), denoted
by $j_{x}^{k}f$, of a smooth map $f:Q\rightarrow N$ at a
point $x\in Q$ (see Figure \ref {jet1}), is defined as an
\emph{equivalence class} of $k-$tangent maps at $x$,
\begin{equation*}
j_{x}^{k}f:Q\rightarrow N=\{f':f'\text{ is }k-\text{tangent to
}f\text{ at }x\}.
\end{equation*}

For example, consider a simple function
~$f:X\rightarrow Y,\,x\mapsto y=f(x)$, mapping the $X-$axis
into the $Y-$axis in $\mathbb{R}^2$. At a chosen point
$x\in X$ we have:\\ a $0-$jet is a graph: $(x,f(x))$;\\
a $1-$jet is a triple: $(x,f(x),f{'}(x))$;\\ a
$2-$jet is a quadruple: $(x,f(x),f{'}(x),f^{\prime \prime }(x))$,\\
~~ and so on, up to the order $k$ (where
$f{'}(x)=\frac{df(x)}{dx}$, etc).\\ The set of all $k-$jets from
$j^k_xf:X\rightarrow Y$ is called the $k-$jet manifold
$J^{k}(X,Y)$.

Formally, given a biomechanical bundle $\Y$, its first order {\it jet manifold} $J^1Y$ comprises the set of
equivalence classes $j^1_xs$, $x\in X$, of sections $s:X\to Y$
so that sections
$s$ and $s'$ belong to the same class iff
$$
Ts\mid _{T_xX} =Ts'\mid_{T_xX}.
$$
Intuitively, sections $s,s'\in j^1_xs$  are identified by their values
$s^i(x)={s'}^i(x)$ and the values of their partial derivatives
$\dr_\mu s^i(x)=\dr_\mu{s'}^i(x)$
at the point $x$ of $X$. There are the natural fibrations \cite{book}
$$
\pi_1:J^1Y\ni j^1_xs\mapsto x\in X, \qquad
\pi_{01}:J^1Y\ni j^1_xs\mapsto s(x)\in Y.
$$
Given bundle coordinates $(x^\la,y^i)$ of $Y$, the associated jet manifold $J^1Y$
is endowed with the adapted coordinates
\be
(x^\la,y^i,y_\la^i), \qquad (y^i,y_\la^i)(j^1_xs)=(s^i(x),\dr_\la
s^i(x)), \qquad
{y'}^i_\la = \frac{\dr x^\m}{\dr{x'}^\la}(\dr_\m +y^j_\m\dr_j)y'^i.
\ee

In particular, given the biomechanical configuration bundle $Q\rightarrow \mathbb{R}$ over
the time axis $\mathbb{R}$, the \textit{$1-$jet
space} $J^{1}(\mathbb{R},Q)$
is the set of equivalence classes $j_{t}^{1}s$ of sections $s^{i}:\mathbb{R}%
\rightarrow Q$ of the configuration bundle $Q\rightarrow \mathbb{R}$, which are
identified by their values $s^{i}(t)$, as well as by the values of their partial derivatives $%
\partial _{t}s^{i}=\partial _{t}s^{i}(t)$ at time points $t\in \mathbb{R}$.
The 1--jet manifold $J^{1}(\mathbb{R},Q)$ is coordinated by $(t,q^{i},\dot{q}%
^{i})$, that is by \textsl{(time, coordinates and velocities)} at every active human joint, so the 1--jets are local joint coordinate maps
\begin{equation*}j_{t}^{1}s:\mathbb{R}%
\rightarrow Q,\qquad t\mapsto (t,q^{i},\dot{q}^{i}).
\end{equation*}

The {\it repeated jet manifold}
$J^1J^1Y$  is defined to be the jet manifold of the bundle
$J^1Y\to X$. It
is endowed with the adapted coordinates $(x^\la ,y^i,y^i_\la
,y_{(\m)}^i,y^i_{\la\m})$.

The {\it second order jet manifold} $J^2Y$
of a bundle $\Y$ is the subbundle of $\wh J^2Y\to J^1Y$ defined
by the coordinate conditions
$y^i_{\la\m}=y^i_{\m\la}$. It has the local coordinates
$(x^\la ,y^i, y^i_\la,y^i_{\la\leq\m})$
together with the transition functions \cite{book}
\be
{y'}_{\la\m}^i= \frac{\dr x^\al}{\dr{x'}^\m}(\dr_\al +y^j_\al\dr_j
+y^j_{\nu\al}\dr^\nu_j){y'}^i_\la.
\ee
The second order jet manifold $J^2Y$  of $Y$ comprises
the equivalence classes  $j_x^2s$ of sections $s$ of\\ $Y\to X$ such that
\be
y^i_\la (j_x^2s)=\dr_\la s^i(x),\qquad
y^i_{\la\m}(j_x^2s)=\dr_\m\dr_\la s^i(x).
\ee
In other words, two sections $s,s'\in j^2_xs$ are identified by their values
and the values of their first and second order
derivatives at the point $x\in X$.

In particular, given the biomechanical configuration bundle $Q\rightarrow \mathbb{R}$ over
the time axis $\mathbb{R}$, the \textit{$2-$jet space} $J^{2}(\mathbb{R},Q)$
is the set of equivalence classes $j_{t}^{2}s$ of sections $s^{i}:\mathbb{R}\rightarrow Q$%
\ of the configuration bundle $\pi:Q\rightarrow \mathbb{R}$, which
are identified by their values $s^{i}(t)$, as well as the values
of their first and second partial derivatives, $\partial
_{t}s^{i}=\partial _{t}s^{i}(t)$
and $\partial _{tt}s^{i}=\partial _{tt}s^{i}(t)$, respectively, at time points $%
t\in \mathbb{R}$. The 2--jet manifold $J^{2}(\mathbb{R},Q)$ is
coordinated by $(t,q^{i},\dot{q}^{i},\ddot{q}^{i})$, that is by \textsl{(time, coordinates, velocities and accelerations)} at every active human joint, so the
2--jets are local joint coordinate maps\footnote{For more technical details on jet spaces with their physical applications, see \cite{Saunders,massa,book,sard98}).}
\begin{equation*}j_{t}^{2}s:\mathbb{R}%
\rightarrow Q,\qquad t\mapsto
(t,q^{i},\dot{q}^{i},\ddot{q}^{i}).
\end{equation*}

\section{Lagrangian Time--Dependent Biomechanics}

\subsection{Jet Dynamics and Quadratic Equations}

The general form of time-dependent Lagrangian biomechanics with \emph{time-dependent
Lagrangian} function $L(t;q^{i};\dot{q}^{i})$ defined on the {jet
space} $X=J^{1}(\mathbb{R},Q)\cong
\mathbb{R}\times TQ$, with local canonical coordinates:
$(t;q^{i};\dot{q}^{i})=$ (time, coordinates and velocities) in active local joints, can be formulated as \cite{GaneshSprBig,GaneshADG}
\begin{equation}
\frac{d}{dt}L_{\dot{q}^{i}}-L_{q^{i}}=\mathcal{F}_{i}\left( t,q,\dot{q}%
\right) ,\qquad (i=1,...,n),  \label{classic}
\end{equation}%
where the coordinate and velocity partial derivatives of the Lagrangian are
respectively denoted by $L_{q^{i}}$ and $L_{\dot{q}^{i}}$.

The most interesting instances of (\ref{classic}) are quadratic biomechanical
equations, of the general form
\begin{equation}
\xi ^{i}\equiv\ddot{q}^{i}=a_{jk}^{i}(q^{\mu })\dot{q}^{j}\dot{q}^{k}+b_{j}^{i}(q^{\mu
})\dot{q}^{j}+f^{i}(q^{\mu }).  \label{cqg100}
\end{equation}%
They are coordinate--independent due to the affine
transformation law of coordinates $\dot{q}^{i}$. Then, it is clear
that the corresponding dynamical connection $\G _{\xi }$
is affine \cite{GaneshSprBig,GaneshADG}:
\[
\G _{\xi } =dq^{\alpha }\otimes \lbrack \partial _{\alpha }+(\G
_{\lambda 0}^{i}(q^{\nu })+\G _{\lambda j}^{i}(q^{\nu
})\dot{q}^{j})\partial _{i}^{t}].
\]%
This connection is symmetric iff $\G
_{\lambda \mu }^{i}=\G _{\mu \lambda }^{i}$.

There is 1--1 correspondence between the affine connections
$\G $
on the affine jet bundle\\ $J^{1}(\mathbb{R},Q)\rightarrow Q$ and the linear connections $%
K$ on the tangent bundle $TQ\rightarrow
Q$ of the autonomous biomechanical manifold $Q$.
This correspondence is given by the relation
\[
\G _{\mu }^{i}=\G _{\mu 0}^{i}+\G _{\mu
j}^{i}\dot{q}^{j},\qquad \G _{\mu \lambda }^{i}=K_{\mu
}{}^{i}{}_{\alpha }.
\]%

Any quadratic biomechanical equation (\ref{cqg100}) is equivalent to the geodesic equation \cite{book}
$$
\dot{t}=1,\qquad \ddot{t}=0,  \qquad
\ddot{q}^{i}=a_{jk}^{i}(q^{\mu })\dot{q}^{i}\dot{q}^{j}+b_{j}^{i}(q^{\mu })%
\dot{q}^{j}\dot{t}+f^{i}(q^{\mu })\dot{t}\dot{t},
$$
for the symmetric linear connection
\[
K=dq^{\alpha }\otimes (\partial _{\alpha }+K_{\alpha
\nu }^{\mu }{}(t,q^i )\dot{q}^{\nu }\dot{\partial}_{\mu })
\]%
on the tangent bundle $TQ\rightarrow Q$, given by the components
$$
K_{\alpha \nu }^{0}{}=0,\quad K_{0}{}^{i}{}_{0}=f^{i},\quad
K_{0}{}^{i}{}_{j}=K_{j}{}^{i}{}_{0}=\frac{1}{2}b_{j}^{i},\quad
K_{j}{}^{i}{}_{k}=a_{jk}^{i}.
$$
Conversely, any linear connection $K$ on the tangent bundle
$TQ\rightarrow Q$ defines the quadratic dynamical equation
\[
\ddot{q}^{i}=K_{j}{}^{i}{}_{k}\dot{q}^{j}\dot{q}^{k}+(K_{0}{}^{i}{}_{j}+K_{j}{}^{i}{}_{0})\dot{q}^{j}+K_{0}{}^{i}{}_{0},
\]%
written with respect to a given reference frame $(t,q^{i})\equiv
q^{\mu}$ (see \cite{book} for technical details).

\subsection{Local Muscle--Joint Mechanics}

The right--hand side terms $\mathcal{F}_{i}(t,q,\dot{q})$ of
(\ref{classic}) denote any type of {external} torques and forces,
including excitation and contraction dynamics of
muscular--actuators and rotational dynamics of hybrid robot
actuators, as well as (nonlinear) dissipative joint torques and
forces and external stochastic perturbation torques and forces. In
particular, we have
\cite{GaneshSprSml,GaneshWSc}):

\textbf{1. Synovial joint mechanics}, giving the first stabilizing effect
to the conservative skeleton dynamics, is described by the
$(q,\dot{q})$--form of the {Rayleigh--Van der Pol's dissipation
function}
\begin{equation*}
R=\frac{1}{2}\sum_{i=1}^{n}\,(\dot{q}^{i})^{2}\,[\alpha _{i}\,+\,\beta
_{i}(q^{i})^{2}],\quad
\end{equation*}
where $\alpha _{i}$ and $\beta _{i}$ denote dissipation parameters. Its
partial derivatives give rise to the viscous--damping torques and forces in
the joints
\begin{equation*}
\mathcal{F}_{i}^{joint}=\partial R/\partial \dot{q}^{i},
\end{equation*}
which are linear in $\dot{q}^{i}$ and quadratic in $q^{i}$.

\textbf{2. Muscular mechanics}, giving the driving torques and forces $%
\mathcal{F}_{i}^{muscle}=\mathcal{F}_{i}^{muscle}(t,q,\dot{ q})$ with $%
(i=1,\dots ,n)$ for human biomechanics, describes the internal {excitation} and
{contraction} dynamics of {equivalent muscular actuators} \cite%
{Hatze}.

(a) \emph{Excitation dynamics} can be described by an impulse {%
force--time} relation
\begin{eqnarray*}
F_{i}^{imp} &=&F_{i}^{0}(1\,-\,e^{-t/\tau _{i}})\text{ \qquad if stimulation
}>0 \\
\quad F_{i}^{imp} &=&F_{i}^{0}e^{-t/\tau _{i}}\qquad \qquad \;\quad\text{if
stimulation }=0,\quad
\end{eqnarray*}
where $F_{i}^{0}$ denote the maximal isometric muscular torques
and forces, while $\tau _{i}$ denote the associated time
characteristics of particular muscular actuators. This relation
represents a solution of the Wilkie's muscular {active--state
element} equation \cite{Wilkie}
\begin{equation*}
\dot{\mu}\,+\,\G \,\mu \,=\,\G \,S\,A,\quad \mu (0)\,=\,0,\quad
0<S<1,
\end{equation*}
where $\mu =\mu (t)$ represents the active state of the muscle, $\G $
denotes the element gain, $A$ corresponds to the maximum tension the element
can develop, and $S=S(r)$ is the `desired' active state as a function of the
motor unit stimulus rate $r$. This is the basis for biomechanical force controller.

(b) \emph{Contraction dynamics} has classically been described by the
Hill's {hyperbolic force--velocity }relation \cite{Hill}
\begin{equation*}
F_{i}^{Hill}\,=\,\frac{\left( F_{i}^{0}b_{i}\,-\,\delta _{ij}a_{i}\dot{q}%
^{j}\,\right) }{\left( \delta _{ij}\dot{q}^{j}\,+\,b_{i}\right) },\,\quad
\end{equation*}
where $a_{i}$ and $b_{i}$ denote the {Hill's parameters},
corresponding to the energy dissipated during the contraction and
the phosphagenic energy conversion rate, respectively, while
$\delta _{ij}$ is the Kronecker's $\delta-$tensor.

In this way, human biomechanics describes the excitation/contraction dynamics for the $i$th
equivalent muscle--joint actuator, using the simple impulse--hyperbolic product relation
\begin{equation*}
\mathcal{F}_{i}^{muscle}(t,q,\dot{q})=\,F_{i}^{imp}\times F_{i}^{Hill}.\quad
\end{equation*}

Now, for the purpose of biomedical engineering and rehabilitation,
human biomechanics has developed the so--called \emph{hybrid rotational actuator}. It
includes, along with muscular and viscous forces, the D.C. motor
drives, as used in robotics \cite{Vuk,GaneshSprSml}
\begin{eqnarray*}
&&\mathcal{F}_{k}^{robo}=i_{k}(t)-J_{k}\ddot{q}_{k}(t)-B_{k}\dot{q}_{k}(t),\qquad\text{with}\\
&&l_{k}i_{k}(t)+R_{k}i_{k}(t)+C_{k}\dot{q}_{k}(t)=u_{k}(t),
\end{eqnarray*}
where $k=1,\dots,n$, $i_{k}(t)$ and $u_{k}(t)$ denote currents and voltages
in the rotors of the drives, $R_{k},l_{k}$ and $C_{k}$ are resistances,
inductances and capacitances in the rotors, respectively, while $J_{k}$ and $%
B_{k}$ correspond to inertia moments and viscous dampings of the drives,
respectively.

Finally, to make the model more realistic, we need to add some stochastic
torques and forces \cite{NeuFuz}
\begin{equation*}
\mathcal{F}_{i}^{stoch}=B_{ij}[q^{i}(t),t]\,dW^{j}(t),
\end{equation*}
where $B_{ij}[q(t),t]$ represents continuous stochastic {diffusion
fluctuations}, and $W^{j}(t)$ is an $N-$variable {Wiener process}
(i.e., generalized Brownian motion), with
$$dW^{j}(t)=W^{j}(t+dt)-W^{j}(t),\qquad (\text{for} ~~j=1,\dots,N).$$

\section{Time--Dependent Riemannian Geometry of Biomechanics}

As illustrated in the introduction, the mass-inertia matrix of human body, defining the Riemannian metric tensor $g_{ij}(q)$ need not be time-constant, as in case of fast gymnastic movements and pirouettes in ice skating, which are based on quick variations of inertia moments and products constituting the material metric tensor $g_{ij}(q)$. In particular, in the geodesic framework (\ref{geodes}), the (in)stability of the
biomechanical joint and center-of-mass trajectories is the (in)stability of the geodesics, and it is
completely determined by the curvature properties of the
underlying manifold according to the \textit{Jacobi equation} of
\emph{geodesic deviation} \cite{GaneshSprBig,GaneshADG}
\begin{equation*}
\frac{D^{2}J^{i}}{ds^{2}}+R_{~jkm}^{i}\frac{dq^{j}}{ds}J^{k}\frac{dq^{m}}{ds}%
=0,
\end{equation*}%
whose solution $J$, usually called \textit{Jacobi variation
field}, locally measures the distance between nearby geodesics;
$D/ds$ stands for the \textit{covariant derivative} along a
geodesic and $R_{~jkm}^{i}$ are the components of the
\textit{Riemann curvature tensor}.

In general, the biomechanical metric tensor $g_{ij}$ is both time and joint dependent, $g_{ij}=g_{ij}(t,q)$. This time-dependent Riemannian geometry can be formalized in terms of the
\textit{Ricci flow} \cite{Ham82}, the
nonlinear heat--like evolution metric equation:
\begin{equation}
\partial _{t}g_{ij}=-R_{ij},  \label{RF}
\end{equation}%
for a time--dependent Riemannian metric $g=g_{ij}(t)$ on a smooth $n-$manifold $Q$ with the Ricci curvature tensor $%
R_{ij} $. This equation roughly says
that we can deform any metric on the configuration manifold $Q$ by the
negative of its curvature; after \emph{normalization}, the final state of such deformation will be a metric with constant
curvature. The negative sign in (\ref{RF}) insures a kind of global \emph{volume
exponential decay},\footnote{%
This complex geometric process is globally similar to a generic exponential
decay ODE:
\begin{equation*}
\dot{q}=-\lambda f(q),
\end{equation*}%
for a positive function $f(q)$. We can get some insight into its solution
from the simple exponential decay ODE,
\begin{equation*}
\dot{q}=-\lambda q\qquad \text{with the solution}\qquad q(t)=q_{0}\mathrm{e}%
^{-\lambda t},
\end{equation*}%
where $q=q(t)$ is the observed quantity with its initial value $q_{0}$ and $%
\lambda $ is a positive decay constant.} since the Ricci flow equation (\ref{RF}) is a kind of nonlinear
geometric generalization of the standard linear \emph{heat
equation}
\begin{equation*}
\partial _{t}u=\Delta u.
\end{equation*}

In a suitable local coordinate system, the Ricci flow equation (\ref{RF}) on a biomechanical configuration manifold $Q$
has a nonlinear heat--type form, as follows. At any time $t$, we can choose
local harmonic coordinates so that the coordinate functions are locally
defined harmonic functions in the metric $g(t)$. Then the Ricci flow takes
the general form \cite{GaneshADG}
\begin{eqnarray}
\partial _{t}g_{ij}&=&\Delta _{Q}g_{ij}+G_{ij}(g,\partial g), \qquad\text{where}  \label{RH} \\
\Delta _{Q}&\equiv & \frac{1}{\sqrt{\det (g)}}\frac{\partial }{\partial q^{i}}%
\left( \sqrt{\det (g)}g^{ij}\frac{\partial }{\partial
q^{j}}\right) \notag
\end{eqnarray} is the \textit{Laplace--Beltrami operator} of the configuration manifold $Q$
and $G_{ij}(g,\partial g)$ is a lower--order term
quadratic in $g$ and its first order partial derivatives $\partial
g$. From the analysis of nonlinear heat PDEs, one obtains
existence and uniqueness of forward--time solutions to the Ricci
flow on some time interval, starting at any smooth initial metric
$g_{0}$ on $Q$.

The exponentially-decaying geometrical diffusion (\ref{RH}) is a formal description for pirouettes in ice skating and fast rotational movements in gymnastics.

\section{Conclusion}

We have presented the time-dependent generalization of an `ordinary' autonomous human biomechanics, in which total mechanical + biochemical energy is not conserved. We have introduced a general framework for time-dependent biomechanics in terms of jet manifolds associated to the extended musculo-skeletal configuration manifold, called the configuration bundle. We start with an ordinary, autonomous configuration manifold of human body motion, given as a set of its all active DOF for a particular movement. This is a Riemannian manifold with a material metric tensor given by the total mass-inertia matrix of the human body segments. This is the base manifold for standard autonomous biomechanics. To make its time-dependent generalization, we had to extend it with a real time axis. By this extension, using techniques from fibre bundles, we defined the biomechanical configuration bundle. On the biomechanical bundle we defined vector-fields, differential forms and affine connections, as well as first and second biomechanical jet manifolds of velocities and accelerations and prolongations of locomotion vector-fields. Using the formalism of jet manifolds of velocities and accelerations, we have developed the time-dependent Lagrangian biomechanics. Finally, we haver showed that the underlying geometric evolution is given by the parabolic Ricci flow equation.

\end{document}